\documentclass[%
 reprint,
 amsmath,amssymb, aps, prd,
]{revtex4-2}

\usepackage{graphicx}% Include figure files
\usepackage{dcolumn}% Align table columns on decimal point
\usepackage{bm}% bold math
\usepackage{hyperref}% add hypertext capabilities
\newcommand{\bdm}{\begin{displaymath}}
\newcommand{\edm}{\end{displaymath}}
\newcommand{\beq}{\begin{equation}}
\newcommand{\eeq}{\end{equation}}
\newcommand{\bea}{\begin{eqnarray}}
\newcommand{\eea}{\end{eqnarray}}

\newcommand{\lt}{\left}
\newcommand{\rt}{\right}
\newcommand{\no}{\nonumber}
\newcommand{\nn}{\nonumber\\}
\newcommand{\ov}{\overline}
\newcommand{\eq}[1]{Eq.~(\ref{#1})}
\newcommand{\eqsand}[2]{Eqs.~(\ref{#1}) and (\ref{#2})}
\newcommand{\eqsto}[2]{Eqs.~(\ref{#1}) to (\ref{#2})}
\newcommand{\gev}{\,\mbox{GeV}}
\newcommand{\tev}{\,\mbox{TeV}}
\newcommand{\imag}{\mbox{Im\,}}
\newcommand{\real}{\mbox{Re\,}}

\newcommand{\Bbar}{\bar{B}}

\newcommand{\bbd}{\ensuremath{B_d\!-\!\Bbar{}_d\,}}
\newcommand{\bbs}{\ensuremath{B_s\!-\!\Bbar{}_s\,}}
\newcommand{\bbq}{\ensuremath{B_q\!-\!\Bbar{}_q\,}}
\newcommand{\bbms}{\bbs\ mixing}

\newcommand{\bbmq}{\bbq\ mixing}

\newcommand{\bra}[1]{\ensuremath{\langle #1 |}}
\newcommand{\ket}[1]{\ensuremath{| #1 \rangle }}
\newcommand{\fig}[1]{Fig.~\ref{#1}}
\newcommand{\tab}[1]{Tab.~\ref{#1}}
\newcommand{\lqcd}{\Lambda_{\rm QCD}} 

\newcommand{\dm}{\ensuremath{\Delta M}}
\newcommand{\dg}{\ensuremath{\Delta \Gamma}}
\newcommand{\epm}[2]{
 \raisebox{-0.5ex}{\shortstack[l]{$\scriptstyle+#1$\\$\scriptstyle-#2$}}}

\newcommand{\be}{\begin{equation}}
\newcommand{\ee}{\end{equation}}

\begin{document}

%\preprint{TTP-20-026, P3H-20-028}

\boldmath
\title{~\hfill \mbox{\normalsize\normalfont TTP-20-026, P3H-20-028}\\[2mm]
  Penguin contribution to width difference and CP asymmetry
  in $B_q\!-\!\Bbar{}_q\,$ mixing at order $\alpha_s^2 N_f$\\ }
\unboldmath

\author{Hrachia~M.~Asatrian$^{1}$}  \email{hrachia@yerphi.am}
\author{Hrachya~H.~Asatryan$^{2}$}	\email{hrachasatryan48@gmail.com}
\author{Artyom~Hovhannisyan$^{1}$}  \email{artyom@yerphi.am}
\author{Ulrich~Nierste$^{3}$}       \email{ulrich.nierste@kit.edu}
\author{Sergey~Tumasyan$^{2}$}      \email{sergey.tumasyan@gmail.com}
\author{Arsen~Yeghiazaryan$^{1}$}   \email{arsen.yeghiazaryan@gmail.com}

\affiliation{~\\
$^1$ Yerevan Physics Institute, 0036 Yerevan, Armenia
\\
$^2$ Yerevan State University, 0025 Yerevan, Armenia
\\
$^3$ Institut f{\"u}r Theoretische Teilchenphysik, Karlsruher
    Institut f{\"u}r Technologie, 76131 Karlsruhe, Germany }

%\date{\today}

\begin{abstract}
  We present new contributions to the decay matrix element
  $\Gamma_{12}^q$ of the $B_q$-$\bar B_q$ mixing complex, where
  $q=d$ or $s$. 
  Our new results  constitute the order $\alpha_s^2 N_f$ corrections
  to the penguin contributions to the Wilson coefficients entering
  $\Gamma_{12}^q$ with full dependence on the charm quark mass.
  This is the first step towards the prediction of the
  CP asymmetry $a_{\rm fs}^q$ quantifying CP violation in mixing  at
  next-to-next-to-leading logarithmic order (NNLO) in quantum chromodynamics
  (QCD) and further improves the prediction of the width difference $\dg_q$
  between the two neutral-meson eigenstates. 
  We find a sizable effect from the non-zero charm mass and our partial
  NNLO result decreases  the NLO
  penguin corrections to $a_{\rm fs}^q$ by 37\% and
  those to $\dg_q$ by 16\%. 
  We further update the Standard-Model 
  NLO predictions for $a_{\rm fs}^q$ and the ratio of the width and mass
  differences of the $B_q$ eigenstates: If we express the results in
  terms of the pole mass of the bottom quark we find   
    $a^s_{\rm fs} =(2.07 \pm 0.10)\cdot 10^{-5}$,
  $a^d_{\rm fs} = (-4.71 \pm 0.24)\cdot 10^{-4}$,
$\Delta{\Gamma}_s/\Delta{M}_s = (4.33 \pm 1.26)\cdot 10^{-3}$,
and  $\Delta{\Gamma}_d/\Delta{M}_d = (4.48 \pm 1.19)\cdot 10^{-3}$.
In the $\overline{\rm MS}$ scheme these numbers read
  $a^s_{\rm fs} =(2.04 \pm 0.11)\cdot 10^{-5}$, 
  $a^d_{\rm fs} = (-4.64 \pm 0.25)\cdot 10^{-4}$,
    $\Delta{\Gamma}_s/\Delta{M}_s = (4.97 \pm 1.02)\cdot 10^{-3}$, and
  $\Delta{\Gamma}_d/\Delta{M}_d = (5.07 \pm 0.96)\cdot 10^{-3}$.
\end{abstract}

\maketitle

\section{\label{sec:level1}Introduction}
Flavor-changing neutral current (FCNC) processes probe new physics
with masses far beyond the reach of future particle colliders. This
justifies the experimental effort at dedicated experiments like LHCb
\cite{Alves:2008zz} and Belle II \cite{Kou:2018nap}.  The \bbd\ and
\bbms\ amplitudes are sensitive to tree-level exchanges of potential new
particles with masses above 100\tev. The oscillations between the flavor
eigenstates $B_q$ and $\Bbar_q$, where $q=d$ or $s$, are governed by two
$2\times 2$ matrices, the mass matrix $M$ and the decay matrix
$\Gamma$. The inclusive, i.e.\ process-independent quantities entering
all oscillation phenomena are $|M_{12}^q|$, $|\Gamma_{12}^q|$, and
$\arg(-M_{12}^q/\Gamma_{12})^q$. Diagonalizing $M^q-i\Gamma^q/2$ gives
the mass eigenstates $B^q_L$ and $B^q_H$ with the subscripts denoting
``light'' and ``heavy'', respectively. The eigenvalues
$M^q_L-i \Gamma^q_L/2$ and $M^q_H-i \Gamma_H^q/2$ define masses and
decay widths of $B^q_L$ and $B^q_H$, which obey exponential decay laws.
The above-mentioned three fundamental physical quantities of \bbmq\ can
be found by measuring $\dm_q = M^q_H-M^q_L $ (coinciding with the \bbmq\
oscillation frequency), $\dg_q=\Gamma^q_L-\Gamma^q_H$, and  \cite{hw}
\begin{eqnarray}
  a^q_{\rm fs}&= & \imag \frac{\Gamma^q_{12}}{M^q_{12}} .
                   \label{defafs}
\end{eqnarray}
The standard way to measure $ a^q_{\rm fs}$ involves the semileptonic
CP asymmetry
\begin{eqnarray}
a^q_{\rm sl}&=&\frac{\Gamma\left(\Bbar_q(t)\to X \ell^+ \nu_\ell
  \right)-\Gamma\left(B_q (t)\to  \bar X \ell^- \bar\nu_\ell  \right)}
  {\Gamma\left(\Bbar_q(t)\to X \ell^+ \nu_\ell
  \right)+\Gamma\left(B_q (t)\to  \bar X \ell^- \bar\nu_\ell  \right)} .
\end{eqnarray}
In the absence of direct CP violation in the semileptonic decay
amplitude one has $ a^q_{\rm fs}=a^q_{\rm sl}$. Direct CP
violation in $B \to X \ell^+ \nu_\ell$ is extremely suppressed in the
Standard Model (SM), so that this identification is justified.
(In all plausible models of new physics
this statement holds as well for $B \to D \ell^+ \nu_\ell$, because
the needed CP-conserving phase comes from QED corrections only.)
The ratio  $\dg_q/\dm_q$ is given by
\begin{eqnarray}
  \frac{\dg_q}{\dm_q}&= & -\real \frac{\Gamma^q_{12}}{M^q_{12}} .
                          \label{defdm}
\end{eqnarray}
In this paper we report on new contributions
to  $\Gamma^q_{12}/M^q_{12}$ which constitute a portion of
the next-to-next-to-leading order (NNLO) QCD corrections to
$ a^q_{\rm fs}$ in \eq{defafs} and  $\dg_q/\dm_q$ in \eq{defdm}.

The mass differences $\dm_s = (17.757 \pm 0.021)\, \mbox{ps}^{-1}$ and
$\dm_d = (0.5064 \pm 0.0019)\, \mbox{ps}^{-1}$ \cite{hfag,Amhis:2016xyh}
have been determined very precisely by the CDF \cite{Abulencia:2006ze}
and LHCb \cite{Aaij:2013mpa} experiments from the \bbq\ oscillation
frequencies. The experimental values of the width differences
\cite{hfag,Amhis:2016xyh},
\begin{eqnarray}
\dg^{\rm exp}_s  &=& \quad\; (8.9 \pm 0.6)\cdot 10^{-2}\,\, \mbox{ps}^{-1}
\\
\dg^{\rm exp}_d  &=& (-1.32 \pm 6.58)\cdot 10^{-3}\,\, \mbox{ps}^{-1}
\label{eq:dgexp}
\end{eqnarray}
are based on measurements by LHCb \cite{lhcb,lhcb2}, ATLAS
\cite{Aad:2016tdj}, CMS \cite{Khachatryan:2015nza}, and CDF
\cite{Aaltonen:2012ie}.  The current experimental world averages for the
semileptonic asymmetries are \cite{hfag,Amhis:2016xyh}
\begin{align}
a_{\rm sl}^{s\rm ,exp}  &= \;(60 \pm 280)\cdot 10^{-5}\,,
\\
a_{\rm sl}^{d\rm ,exp}  &= (-21 \pm 17)\cdot 10^{-4}.
\label{eq:afsexp}
\end{align}
Clearly, $\dg_s$ is a precision observable, while the three other
quantities are still far from giving precise information on
fundamental parameters. For $a_{\rm fs}^d$ and $\dg_d$ it is worthwhile
to study the clean sample of $B\to J/\psi K_s$ decays \cite{Nierste:2004uz}. 
While new physics will primarily enter $M_{12}^q$, 
scenarios in which $\Gamma_{12}^q$ is affected have been studied as well
\cite{Jager:2017gal,Bobeth:2014rda}, especially the doubly
Cabibbo-suppressed  $\Gamma_{12}^d$ could play a role in new-physics
studies.

The state-of-the-art of the theory predictions of $a_{\rm fs}^d$ and
$\dg_d$ is next-to-leading logarithmic order (NLO) QCD for the leading-power
contribution \cite{NiersteNLO,Ciuchini:2003ww,Beneke:2003az,NiersteNLONB}
and LO QCD for the ${\cal O} (\lqcd/m_b)$ power-suppressed
corrections \cite{Beneke:1996gn,NiersteNLONB}. The accuracy of $\dg^{\rm
  exp}_s$ in \eq{eq:dgexp} calls for an NNLO
calculation, which is a 
formidable project. First steps in this direction have been made in
Ref.~\cite{Asatrian:2017qaz}, in which terms of order
$\alpha_s^2 N_f$ to $\Gamma_{12}$, where $N_f=5$ is the number of active quark flavors,
have been calculated up to order $m_c/m_b$.
This calculation has permitted a better assessment of  $\dg_q$, but not of
$a_{\rm fs}^q$, which is proportional to $m_c^2/m_b^2$. 

The purpose of the present paper is to do the next step in the
calculation of NNLO QCD
corrections to $\Gamma_{12}$. We calculate the penguin contributions
with full dependence on the charm quark mass. These terms constitute an
improvement for the prediction of $\dg_q$ compared to
Ref.~\cite{Asatrian:2017qaz}, and, more importantly, are the first step
towards the prediction of  $a^q_{\rm fs}$ at NNLO accuracy. 

Penguin contributions are small in the Standard Model, because the
Wilson coefficients of the corresponding operators are small, of order
0.05 or smaller. However, this makes these coefficients sensitive to
contributions of new physics, which can easily be of the same size
\cite{Keum:1998fd} as the SM coefficients.  Thus in order to study
such effects beyond the SM a precise knowledge of the penguin
contributions to $\Gamma_{12}^q$ is desirable.

This paper is organized as follows: In the following section we
summarize the theoretical framework of the calculation. In Section
\ref{sec:cf} we present our analytical results and subsequently perform a
phenomenological analysis in Section \ref{sec:num}. Finally we
conclude. Results for matrix elements needed for the calculation are
relegated to the appendix.

\begin{figure*}[t]
\includegraphics[width=0.9\textwidth]{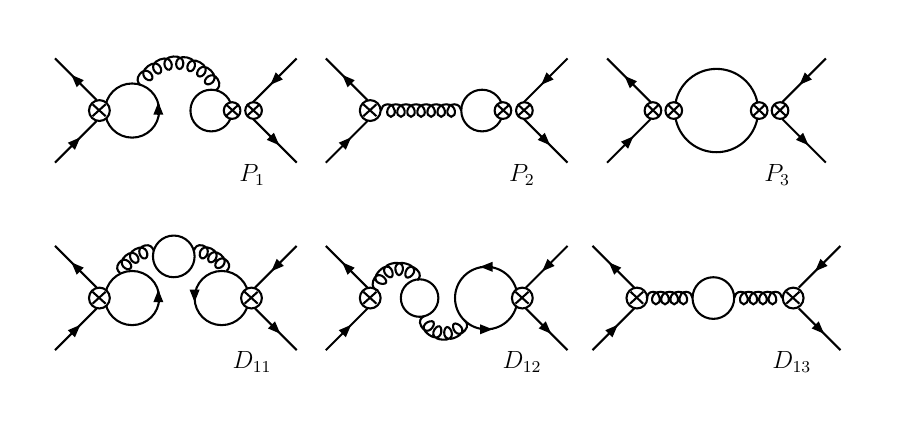}
\vspace{-0.7cm}
\caption{
  Diagrams for the penguin contribution at ${\cal O}(\alpha_s^2 N_f)$.
  The small Wilson coefficients $C_{3-6}$ are counted as
  ${\cal O}(\alpha_s)$. 
  $P_1$, $P_2$ are diagrams with one
  insertion of a penguin operator $O_3,...,O_6$, depicted as two circles
  with crosses, and one insertion of $O_2^{u,c}$ or $O_8$, shown as a single
  circle with cross. $P_3$ 
  denotes a one-loop diagram with two
  insertions of penguin operators $O_3,...,O_6$. $D_{11}$, $D_{12}$ and
  $D_{13}$ are diagrams with insertions of operators
  $O_2^{u,c}$ or $O_8$. (The notation follows
  Ref.~\cite{Asatrian:2017qaz}.)\label{diags}}
~\\[-3mm]
\hrule
\end{figure*}

\section{Theoretical framework\label{sec:tf}}

The effective $\Delta B=1$ weak Hamiltonian, relevant for $b\to s$
transition, reads \cite{Buchalla:1995vs}
\begin{eqnarray}
\label{Heff}
&& H^{\Delta B=1}_{\rm eff}=
\\
&& \nonumber -\frac{G_F}{\sqrt{2}}\left\{\lambda_t^s\left[\sum^6_{i=1}C_iO_i + C_{8}O_{8}\right]
- \lambda_u^s\sum^2_{i=1}C_i\left(O^u_i-O_i\right) \right\}
% \; +\;\mbox{h.c.}
,
\end{eqnarray}
where 
\begin{eqnarray}
%&& \lambda_t^d=V_{td}^*V_{tb},~~\lambda_u^d=V_{ud}^*V_{ub},
%\\
&& \lambda_t^s=V_{ts}^*V_{tb},~~ \lambda_u^s=V_{us}^*V_{ub}
\end{eqnarray}
comprises the elements of the Cabibbo-Kobayashi-Maskawa (CKM) matrix.
The dimension-six effective operators in \eq{Heff} are
\begin{eqnarray}
\label{OperBasis}
\nonumber
O_1^u&=&(\bar{s}_iu_j)_{V-A}\;(\bar{u}_jb_i)_{V-A},~
\hspace{-0.15cm} O_2^u=(\bar{s}_iu_i)_{V-A}\;(\bar{u}_jb_j)_{V-A},
\\
\nonumber
O_1&=&(\bar{s}_ic_j)_{V-A}\;(\bar{c}_jb_i)_{V-A},~
O_2=(\bar{s}_ic_i)_{V-A}\;(\bar{c}_jb_j)_{V-A},
\\
\nonumber
O_3&=&(\bar{s}_ib_i)_{V-A}\;(\bar{q}_jq_j)_{V-A},~
O_4=(\bar{s}_ib_j)_{V-A}\;(\bar{q}_jq_i)_{V-A},
\\
\nonumber
O_5&=&(\bar{s}_ib_i)_{V-A}\;(\bar{q}_jq_j)_{V+A},~
O_6=(\bar{s}_ib_j)_{V-A}\;(\bar{q}_jq_i)_{V+A},
\\
%\nonumber
&& O_{8}=\frac{g_s}{8\pi^2}m_b\bar{s}_i\sigma^{\mu\nu}(1-\gamma_5)T_{ij}^a
b_jG^a_{\mu\nu}.~~~~~~~~~~~~~~~~~
\end{eqnarray}
Here $i,j$ are color indices and summation over $q = u, d, s, c, b$ is
understood. $V\pm A$ denote $\gamma_{\mu}(1\pm \gamma_5)$ and $S\pm P$
(needed below) represents $(1\pm \gamma_5)$.  $C_1,\ldots,C_6$ and $C_8$
are the corresponding Wilson coefficients, which are functions of the
top mass $m_t$ and the $W$ mass $M_W$. $G_F$ is the Fermi constant. The
corresponding formulae for $b\to d$ transitions can be obtained from
Eqs. (\ref{Heff})-(\ref{OperBasis}) by replacing 
$s$ with $d$.

To find $\dg \simeq 2|\Gamma_{12}|$ we must calculate
\begin{eqnarray}
\Gamma_{12} &=& \mbox{Abs}
 \bra{B_s} \,i\!\!\int \!\! d^4x\ T\,{\cal H}^{\Delta B=1}_{\rm eff}(x){\cal H}^{\Delta B=1}_{\rm eff}(0)
 \ket{\Bbar_s}, \quad\label{eq:full}
\end{eqnarray}
where `Abs' denotes the absorptive part of the matrix element and $T$ is
the time ordering operator. Following \cite{Beneke:2003az} we write
$\Gamma_{12}$ as
\begin{eqnarray}
  \Gamma_{12} &=&
                  -\left[\lambda^2_c\Gamma_{12}^{cc}+2\lambda_c\lambda_u\Gamma_{12}^{uc}
                  +\lambda^2_u\Gamma_{12}^{uu}\right]
                  \nn
  &=& -\lambda^2_t\left[ \Gamma_{12}^{cc}\, +\, 2\frac{\lambda_u}{\lambda_t}
     \left(\Gamma_{12}^{cc}-\Gamma_{12}^{uc}\right) \rt. \nn
  &&  \lt. \qquad \; +\,
     \frac{\lambda^2_u}{\lambda_t^2}\left(\Gamma_{12}^{uu}  +
     \Gamma_{12}^{cc}-2\Gamma_{12}^{uc}\right)\right],
\end{eqnarray}
where the coefficients $\Gamma_{12}^{ab}$, $a,b=u,c$ are positive. The
Heavy Quark Expansion (HQE) expresses \eq{eq:full} in terms of matrix
elements of local operators. The leading term (in powers of $\lqcd/m_b$)
reads
\begin{eqnarray}
\hspace{-0.4cm} \Gamma_{12}^{ab} = \frac{G_F^2 m_b^2}{24 \pi\, M_{B_s}}
    \lt[  G^{ab}  \bra{B_s } Q \ket{\Bbar_s} -
          G^{ab}_S \bra{B_s} Q_S \ket{\Bbar_s}
    \rt].  %% \; +\; \Gamma_{12,1/m_b}^{ab}
\label{defg}
\end{eqnarray}
The two $|\Delta B|=2$ operators ($B$ denotes the beauty quantum number) are
\begin{eqnarray}
 &&  Q=(\bar{s}_ib_i)_{V-A}\;(\bar{s}_jb_j)_{V-A},~
 \\
 && \tilde{Q}_S=(\bar{s}_ib_j)_{S-P}\;(\bar{s}_jb_i)_{S-P}. \label{eq:defops}
\end{eqnarray}
The hadronic matrix elements, which are calculated with non-perturbative
methods like lattice QCD, are usually expressed in term of the ``bag''
parameters $ B_{B_q}$, $ B_{S, B_q}^\prime$ as
\begin{eqnarray}
\nonumber \bra{B_q} Q (\mu_2) \ket{\ov B_q}  &=&
   \frac{8}{3} M^2_{B_q}\, f^2_{B_q} B_{B_q}(\mu_2),
\\
\bra{B_q} \widetilde Q_S (\mu_2)\ket{\ov B_q} &=& \frac{1}{3}  M^2_{B_q}\,
  f^2_{B_q} \widetilde B_{S, B_q}^\prime (\mu_2).
      \label{eq:defb}
\end{eqnarray}
Here $f_{B_q}$ is the $B_q$ decay constant and $\mu_2={\cal O}(m_b)$
is the renormalization scale at which the matrix elements are
calculated. In a lattice-QCD calculation $\mu_2$ is the
scale of lattice-continuum matching. In the
expression for $\Gamma_{12}$ the matrix elements of \eq{eq:defb} are
multiplied by perturbative Wilson coefficients depending on
$\mu_2$ as well, resulting in a cancellation of  the unphysical scale $\mu_2$
from $\Gamma_{12}$. Analogously, the dependence on the
renormalization scheme cancels between the  Wilson coefficients and
$B(\mu_2)$, $\widetilde B_S^\prime (\mu_2)$. In this paper we use the
renormalization scheme of Ref.~\cite{NiersteNLO}.

Using the notation of Refs.~\cite{NiersteNLO,Beneke:2003az,
  NiersteNLONB}, we decompose $G^{ab}$ and $G^{ab}_S$ further as
\begin{eqnarray}
G^{ab} \;=\; F^{ab} + P^{ab}, \qquad
G^{ab}_S \; = \; - F^{ab}_S -  P^{ab}_S
    . \label{deffp}
\end{eqnarray}
Here $F^{ab}$ and $F^{ab}_S$ are the contributions from the current-current
operators $O_{1,2}$, while $P^{ab}$ and $P^{ab}_S$ stem
from the penguin operators $O_{3-6}$ and $O_8$. The coefficients
$G^{ab}$, $G^{ab}_S$ are found by applying an operator product expansion
(resulting in the HQE)  to the bilocal matrix
elements (``full theory'')
\begin{eqnarray}
  \mbox{Abs}\,
 \langle \;i\!\int d^4x\ T\, O_i (x) O_j(0) \, \rangle. \label{eq:fab}
\end{eqnarray}
The HQE expresses these bilocal matrix elements in terms of the local
matrix elements $\langle Q \rangle$, $\langle Q_S \rangle$ (``effective
theory''), the coefficients of the latter are the perturbative
short-distance objects studied in this paper.  This matching calculation
can be done order-by-order in the strong coupling $\alpha_s$, with
quarks instead of mesons in the external states in \eq{eq:fab}.  The NLO
result of
Refs.~\cite{NiersteNLO,Beneke:2003az,Ciuchini:2003ww,NiersteNLONB}
contains the result of \eq{eq:fab} at the two-loop level for $i,j=1,2$.
The chromomagnetic operator $O_8$ is proportional to the strong coupling
$g_s$, so that for $i=8$ or $j=8$ NLO accuracy means one loop only.  One
further counts the small penguin Wilson coefficients $C_{3-6}$ as
${\cal O}(\alpha_s)$ and considers only one-loop diagrams for $i\geq 3$
or $j\geq 3$.

\boldmath
\section{Results for the penguin coefficients $P$, $P_S$ at order
  $\alpha_s^2 N_f$ \label{sec:cf}}
\unboldmath

For the contributions of penguin diagrams and penguin operators
in \eq{deffp} we write
\begin{eqnarray}
\nonumber P^{ab}(z) &=& P^{ab,(1)} (z) + P^{ab,(2)}(z),
\\
P^{ab}_S(z) &=& P_S^{ab,(1)}(z) + P_S^{ab,(2)} (z),
\end{eqnarray}
where $P^{ab,(1)}(z)$ and $P_S^{ab,(1)}(z)$ denote the NLO results of
Ref.~\cite{NiersteNLO}, while $P^{ab,(2)}(z)$ and $P_S^{ab,(2)}(z)$ are the
NNLO corrections studied in this paper. Since we treat $C_{3-6}$ as ${\cal
  O}(\alpha_s)$, $P^{ab,(2)}_{(S)}(z)$ contain terms of order $C_{3-6}C_{3-6}$,
$\alpha_s C_2C_{3-6}$, and  $\alpha_s^2C_2^2$. The
large-$N_f$ part of $P^{ab,(2)}(z)$  is decomposed as
\begin{eqnarray}
\nonumber && P^{ab,(2),N_f}(z) = N_H P^{ab,(2),N_H}(1,z)
\\
&& ~~ + N_V P^{ab,(2),N_V}(z_i,z)+N_L P^{ab,(2),N_L}(0,z)
\end{eqnarray}
with an analogous formula for $P_S^{ab,(2)}(z)$. Here, $N_H=1$, $N_V=1$ and $N_L=3$ denote the number of heavy ($b$-quark),
intermediate-mass ($c$-quark) and light $(u, d, s)$ quark flavors, with the total number of quark flavors $N_f = N_H + N_V + N_L =5$. In the penguin
contributions, as well as in charm loops, we keep the charm mass
non-zero, i.e.\ equal to its physical value. This improves our results over
those in Ref.~\cite{Asatrian:2017qaz}, where the charm mass on all lines
touching $O_2$ was set to zero. This affects all loops in the
diagrams in \fig{diags} (see also Figure 1 of
\cite{Asatrian:2017qaz}). The diagrams $P_{1-2}$ are not only needed for
the contributions involving $C_{3-6,8}$, but also appear in counter-term
contributions to $D_{11-13}$, in which the charm mass must be treated in
the same way as in the diagrams which they renormalize.

We introduce the abbreviation $z_i\equiv m_i^2/m_b^2$, where $m_i$
denotes the quark in all closed fermion loops, in which all $N_f=5$
quarks can run.  Thus $z_i=1$, $z_i=m_c^2/m_b^2$, or $z_i=0$ in
$P^{{ab,(2)},N_{H}}(z_i,z)$, $P^{{ab,(2)},N_{V}}(z_i,z)$, or
$P^{{ab,(2)},N_{L}}(z_i,z)$, respectively.  The second argument
$z=m_c^2/m_b^2$ of the loop functions involves the charm mass
originating from $O_{1,2}$ operators.

Our results are:
\begin{eqnarray}
\nonumber P^{cc,(2),N_H}(1,z) &=& \frac{\alpha_s(\mu_1)}{4\pi}
G_p^{cc,(1),N_H}(1,z) M_4'(\mu_1)
\\
&& \hspace{-0.5cm} + \frac{\alpha_s^2(\mu_1)}{(4\pi)^2}
G_p^{cc,(2),N_H}(1,z) C_2^2(\mu_1),
\end{eqnarray}
\begin{eqnarray}
\nonumber P_S^{cc,(2),N_H}(1,z) &=& -\frac{\alpha_s(\mu_1)}{4\pi} 8
G_p^{cc,(1),N_H}(1,z) M_4'(\mu_1)
\\
&& \hspace{-0.5cm} - \frac{\alpha_s^2(\mu_1)}{(4\pi)^2} 8
G_p^{cc,(2),N_H}(1,z) C_2^2(\mu_1),
\end{eqnarray}
\begin{eqnarray}
\nonumber P^{cc,(2),N_V}(z_i,z) &=& \sqrt{1-4z_i}\Big( (1-z_i)M_1'(\mu_1)
\\
                                && \nonumber  \hspace{-1cm}
                                   +
\frac{1}{2}(1-4z_i)M_2'(\mu_1)+3z_iM_3'(\mu_1) \Big)
\\
&&  %% \hspace{3cm}
\nonumber \hspace{-1cm}+ \frac{\alpha_s(\mu_1)}{4\pi}
G_p^{cc,(1),N_V}(z_i,z) M_4'(\mu_1)
\\
&& \hspace{-1cm}+ \frac{\alpha_s^2(\mu_1)}{(4\pi)^2}
G_p^{cc,(2),N_V}(z_i,z) C_2^2(\mu_1), \label{pengz1}
\end{eqnarray}
\begin{eqnarray}
\nonumber && P_S^{cc,(2),N_V}(z_i,z) =
\sqrt{1-4z_i}\,(1+2z_i)\left(M_1'(\mu_1)\right.
\\
&& \nonumber \qquad \left. -M_2'(\mu_1)\right)
- \frac{\alpha_s(\mu_1)}{4\pi}
8G_p^{cc,(1),N_V}(z_i,z) M_4'(\mu_1) 
\\
&& \qquad \qquad- \frac{\alpha_s^2(\mu_1)}{(4\pi)^2}
8G_p^{cc,(2),N_V}(z_i,z) C_2^2(\mu_1),\label{pengz2}
\end{eqnarray}
with
\begin{eqnarray}
\label{NLONH} \nonumber G_p^{cc,(1),N_H}(1,z) &=& -\frac{1}{54}\left(6\log \left(\frac{\mu_1}{m_b}\right) -
3\sqrt{3}\pi+17\right)
\\
&& \times \sqrt{1-4 z} (2 z+1),
\end{eqnarray}
\begin{eqnarray}
\nonumber G_{p}^{cc,(2),N_H}(1,z) &=& \frac{2}{81} \left(6 \log \left(\frac{\mu_1}{m_b}\right) - 3\sqrt{3} \pi +17\right)
\\
&& \nonumber \hspace{-2.0cm} \times \sqrt{1-4 z} (2 z+1)\left[2\log \left(\frac{\mu_1}{m_b}\right)+\frac{2}{3}+4 z\right.
\\
    && \left. \hspace{-2.5cm} -\log (z)+\sqrt{1-4 z} (2 z+1) \log (\sigma)+
                                     \frac{3C_8(\mu_1)}{C_2(\mu_1)}\right],
\end{eqnarray}
\begin{eqnarray}
&& \label{NLONV} \nonumber G_p^{cc,(1),N_V}(z_i,z) =  - \frac{1}{54}
\Big[\sqrt{1-4z_i}(1+2z_i) 
\\
&& \nonumber \hspace{1cm}\qquad  \times \left(6\log\left(\frac{\mu_1}{m_b}\right)-3\log (z)+2+12z\right)
\\
  && \nonumber \hspace{0.5cm}\qquad +\sqrt{1-4z}(1+2z) \nn
        && \hspace{1cm}\qquad
 \times \left(6\log\left(\frac{\mu_1}{m_b}\right)-3\log (z_i)+5+12z_i\right)
\nn
&& \nonumber \hspace{0.5cm}\qquad +
   3\sqrt{1-4z}(1+2z)\sqrt{1-4z_i}(1+2z_i) \nn
&& \hspace{1cm}\qquad   \times \left(\log(\sigma)+\log(\sigma_i)\right)
\nn
&&  \hspace{0.5cm}\qquad + \frac{9C_8(\mu_1)}{C_2(\mu_1)}
\sqrt{1-4z_i}\,(1+2z_i) \Big],
\end{eqnarray}
\begin{eqnarray} \label{pengc}
  && \nonumber G_{p}^{cc,(2),N_V}(z_i,z) =\nn
  && \no    
 \frac{1}{81}\Big\{ 2\sqrt{1-4z}(2z+1)\sqrt{1-4z_i}
 \\
  && \nonumber
     \quad \times (2z_i+1)(\log(\sigma)+\log(\sigma_i)) \nn
&& \no \quad \times     \Big(6\log \Big(\frac{\mu_1}{m_b}\Big)+12z
  -3\log(z)+2\Big) \nn
  && \no + \frac{2}{3}\sqrt{1-4z}(2z+1) \nn
     &&\no \quad\times \Big(6\log\Big(\frac{\mu_1}{m_b}\Big)+5
  +12z_i-3\log(z_i)\Big) \nn
&&\no \quad  \times \Big(6\log\Big(\frac{\mu_1}{m_b}\Big)+2+12z-3\log(z)
 \\
 && \nonumber  \qquad +3\sqrt{1-4z}(2z+1)\log(\sigma)\Big) \nn
&&\no + \frac{1}{3}\sqrt{1-4z_i}(2z_i+1)
\\
&& \nonumber \quad \times \Big[\Big(6\log\Big(\frac{\mu_1}{m_b}\Big)+2+12z-3\log(z)\Big)^2
   \nn
   &&\no \qquad + 9(1-4z)
      (2z+1)^2 \nn
  &&\no
     \qquad \quad\times \Big(2\log(\sigma)\log(\sigma_i)+\log^2(\sigma)-\pi^2\Big)\Big]
\\
  && \nonumber \hspace{0.0cm} +
     \frac{6C_8(\mu_1)}{C_2(\mu_1)}\Big[\sqrt{1-4z}(2z+1)\nn
     &&\no \qquad \times \Big(6\log\Big(\frac{\mu_1}{m_b}\Big)-3\log(z_i)
      +5+12z_i\Big) \nn
  &&\no  \quad + \sqrt{1-4z_i}(2z_i+1)\nn
 &&\no    \qquad \times \Big(6\log\Big(\frac{\mu_1}{m_b}\Big)+2
 +12z-3\log(z)\Big) \nn
&&\no \quad + 3\sqrt{1-4z}(2z+1)\sqrt{1-4z_i}(2z_i+1)
\\
  && \nonumber \qquad\times  \Big(\log(\sigma)+\log(\sigma_i)\Big) \nn
     &&
+ \frac{9C_8(\mu_1)}{2C_2(\mu_1)}\sqrt{1-4z_i}(2z_i+1)\Big]\Big\},
\end{eqnarray}
where we have defined
\begin{eqnarray}
M_1^\prime &=&3C_3^2+2C_3C_4+3C_5^2+2C_5C_6 \nn
         M_2^\prime &=& C_4^2+C_6^2,\nn
                 M_3^\prime& =& 2(3C_3C_5+C_3C_6+C_4C_5+C_4C_6), \nn
                          M_4^\prime&=& 2(C_2C_4+C_2C_6)
\end{eqnarray}
 and
\begin{eqnarray}\label{sigma}
\sigma=\frac{1-\sqrt{1-4z}}{1+\sqrt{1-4z}},
\end{eqnarray}
while $\sigma_i$ is defined by replacing $z$ with $z_i$ in (\ref{sigma}).
Then $P^{uu,(2),N_A}(z_i,0)=P^{cc,(2),N_A}(z_i,0)$ (with $A=H,V,L$) and
\begin{eqnarray}
\nonumber P^{uc,(2),N_A} (z_i,z) &=& \frac{P^{cc,(2),N_A} (z_i,z)+P^{cc,(2),N_A} (z_i,0)}{2}
\\
&& + \Delta P^{uc,(2),N_A},
\\
\nonumber P^{uc,(2),N_A}_S (z_i,z) &=& \frac{P^{cc,(2),N_A}_S (z_i,z)+P^{cc,(2),N_A}_S (z_i,0)}{2}
\\
&& -8 \Delta P^{uc,(2),N_A},
\end{eqnarray}
where
\begin{eqnarray}
&& \nonumber \Delta P^{uc,(2),N_H} = -\frac{\alpha_s^2(\mu_1)}{(4\pi)^2} C_2^2(\mu_1) \frac{1-\sqrt{1-4 z} (1+2 z)}{81} 
\\
\nonumber && \hspace{1cm} \times \Big(6 \log \Big(\frac{\mu_1}{m_b}\Big)-3 \sqrt{3} \pi +17\Big)
\\
&& \hspace{1cm} \times \left[\log (z)-\sqrt{1-4 z} (1+2 z) \log (\sigma)-4 z\right],
\end{eqnarray}
\begin{eqnarray}
  \nonumber && \Delta P^{uc,(2),N_V} =
               \frac{\alpha_s^2(\mu_1)}{(4\pi)^2} C_2^2(\mu_1) \nn
 && \times \frac{1}{162} \Big\{\Big(1-\sqrt{1-4 z} (1
  +2 z)\Big) \nn 
            && \nonumber \quad
               \times \Big[3 \sqrt{1-4 z_i} (1+2 z_i)
\Big[\pi ^2 \Big(1-\sqrt{1-4 z} (1+2 z)\Big)  
\\
            && \nonumber \qquad\quad  +\Big(\sqrt{1-4 z} (1+2 z)+1\Big) \log^2(\sigma )  \nn
 &&\nonumber     \qquad\quad          +2 (4 z-\log (z)) 
    \log (\sigma )\Big]\nn
            &&\nonumber
               \qquad  - 2 \Big(\log (z)-\sqrt{1-4 z} (1+2 z) \log (\sigma )-4 z\Big)
\\
            && \nonumber
               \qquad\quad
               \times\Big(6 \log \lt(\frac{\mu_1}{m_b}\rt) + 3 \sqrt{1-4 z_i} (1+2 z_i) \log (\sigma_i)
\\
&& \nonumber  \qquad +12 z_i-3 \log (z_i)+5\Big)\Big] \nn
&&  \quad -3 \sqrt{1-4 z_i} (1+2 z_i) \Big(16 z^2
\nn
&&  \qquad +(\log (z)-\log (\sigma )) \, (\log (z)-\log (\sigma )-8
   z)\Big)\!\Big\} .
\end{eqnarray}
$P^{ab,(2),N_L}(0,z)$ can be obtained from the expressions presented above
by setting $z_i$ to 0, i.e.\ $P^{ab,(2),N_L}(0,z)= P^{ab,(2),N_V}(0,z)$.

Taking the limit $z\to 0$ in the results presented in this section 
(with the replacement $z_i\to z$)  reproduces the results
in Eqs.~(4.15)-(4.22) of Ref.~\cite{Asatrian:2017qaz}.

\boldmath
\section{Phenomenology of $\Delta\Gamma_q$ and
  $a_{\rm fs}^q$ \label{sec:num}}
\unboldmath%
In this section we first show the impact of a non-zero charm quark mass
in the $\alpha_s^2N_f$ corrections to $\Delta\Gamma_s$ and $a_{\rm fs}^q$,
which is the novel analytic result of this paper. Subsequently we
present updated predictions for  $\dg_q/\dm_q$ and  $a_{\rm fs}^q$,
reflecting the progress in the determination of hadronic parameters,
quark masses, CKM elements, and other parameters entering these
quantities. 

We may express $\Delta\Gamma_q$ and $a_{\rm fs}^q$ in terms of $m_b$ and
$z=m_c^2/m_b^2$. As shown in Ref.~\cite{Beneke:2002rj}, trading $z$ for
$\bar z= (\bar m_c(\bar m_b)/\bar m_b(\bar m_b))^2$ (with the
appropriate changes in the expressions for the radiative corrections)
resums the $z \log z$ terms to all orders, i.e.\ there are no
$\bar z \log \bar z$ terms. In the numerics presented below
we will always use $\bar z$. This still leaves (at least) two natural possibilities
to define $m_b$, two powers of which appear in the prefactor of
$\Delta\Gamma_s$ and $a_{\rm fs}^q$, namely the $\ov{\rm MS}$ mass
$\bar{m}_b(\bar{m}_b)$ and the pole mass  $m^{\rm pole}_b$. In our numerics
we  use $\bar{m}_b(\bar{m}_b) = (4.18\pm 0.03)$ GeV as
input in  both schemes and  calculate $m^{\rm pole}_b = (4.58 \pm
0.03) $ GeV at NLO and $m^{\rm pole}_b = (4.84\pm 0.03) $ GeV at NNLO. 

In our partial NNLO results we further use the complete NNLO
$\Delta B = 1$ Wilson coefficients $C_1$, $C_2$
\cite{Buras:2006gb,Gorbahn:2004my} and the complete NLO expressions for
$C_{3-6}$, $C_8$ (see Ref.~\cite{Asatrian:2017qaz} for details ).  From
the values of $\sin(2\beta)$ and $R_t$ listed in \tab{tab:inp} we obtain
\begin{eqnarray}
\frac{\lambda_u^d}{\lambda_t^d} &=& (0.0122 \pm 0.0097) - (0.4203 \pm 0.0090) i,
\\
  \frac{\lambda_u^s}{\lambda_t^s} &=&
                                      - (0.00865 \pm 0.00042) \nn
                                      &&
                                      + (0.01832\pm 0.00039) i.
\end{eqnarray}

\begin{table*}[t]
\hrule
\begin{displaymath}
\begin{array}{rll@{~~}rll}
\bar{m}_b(\bar{m}_b)=& (4.18 \pm 0.03)\,\gev
&\mbox{\cite{Agashe:2014kda}} & \bar m_c(\bar m_c)=& (1.2982 \pm
0.0013_{\rm stat} \pm 0.0120_{\rm syst})\,\gev &
          \mbox{\cite{Charles:2004jd,Kuhn:2007vp,Allison:2008xk}} \\
\bar m_s(\bar m_b)=& (0.0786 \pm 0.0006)\, \gev &
           \mbox{\cite{Aoki:2019cca}} &
                               % Bazavov:2016nty}} &
\bar m_t (m_t)= & (165.26\pm 0.11_{\rm stat}  \pm 0.30_{\rm syst})\,
   \gev &  \mbox{\cite{Charles:2004jd}}
\\
m_b^{\rm pow} = & 4.7\, \gev & \mbox{\cite{NiersteNLONB}} & \alpha_s(M_Z)= & 0.1181(11) & \mbox{\cite{Tanabashi:2018oca}}
\\
  M_{B_s} =& 5366.88\, \mbox{MeV} & \mbox{\cite{Tanabashi:2018oca}} &
  M_{B_d} =& 5279.64\, \mbox{MeV}  & \mbox{\cite{Tanabashi:2018oca}}
\\
 B_{B_s} =& 0.813 \pm 0.034 &   \mbox{\cite{Dowdall:2019bea}} &
  B_{B_d} =& 0.806 \pm 0.041 &   \mbox{\cite{Dowdall:2019bea}} 
 \\
\widetilde{B}_{S,B_s}^\prime =& 1.31 \pm 0.09
  &\mbox{\cite{Dowdall:2019bea}} &
                                    \widetilde{B}_{S,B_d}^\prime =& 1.20 \pm 0.09 &
                                \mbox{\cite{Dowdall:2019bea}} 
  \\
    {B}_{R_0}^s =& 1.27 \pm 0.52 & \mbox{\cite{Dowdall:2019bea}}
 & {B}_{R_0}^d =& 1.02 \pm 0.55 & \mbox{\cite{Dowdall:2019bea}}
  \\
  {B}_{\tilde R_2}^s =& 0.89 \pm 0.35 & \mbox{\cite{Davies:2019gnp}}
 & {B}_{\tilde R_2}^d =& {B}_{\tilde R_2}^s & 
  \\
 {B}_{\tilde R_3}^s =& 1.14 \pm 0.39 & \mbox{\cite{Davies:2019gnp}}
 & {B}_{\tilde R_3}^d =& {B}_{\tilde R_3}^s & 
  \\
 {B}_{R_2}^q  =& - {B}_{\tilde R_2}^q &   \mbox{\cite{Beneke:1996gn}} &
 {B}_{R_3}^q  =& \frac57 {B}_{\tilde R_3}^q + \frac27 {B}_{\tilde R_2}^q &   \mbox{\cite{Beneke:1996gn}} 
  \\
  f_{B_s} =& (0.2307 \pm  0.0013) \gev & \mbox{\cite{Bazavov:2017lyh}}
                              & f_{B_d} =& (0.1905 \pm 0.0013) \gev &
                                                                      \mbox{\cite{Bazavov:2017lyh}}
  \\
\sin(2\beta) =& 0.7083\epm{0.0127}{0.0098} &
                                             \mbox{\cite{Charles:2004jd}}
                              &
R_t =& 0.9124\epm{0.0064}{0.0100} &
                                    \mbox{\cite{Charles:2004jd}}
  \\
 |V_{us}|= & 0.22483\epm{0.00025}{0.00006} &
                                                \mbox{\cite{Charles:2004jd}}
                              & & &
\end{array}
\end{displaymath}
\caption{Input parameters used in Sec.~\ref{sec:num}.
  $\bar m_s(\bar m_b)$ is calculated from
  $\bar m_s(2\gev)=0.09344\pm 0.00068\,\gev$ \cite{Aoki:2019cca}.
  The listed values for $ B_{B_q}$ and $\widetilde{B}_{S,B_q}^\prime$
  are found by rescaling the numbers in
    Table V of Ref.~\cite{Dowdall:2019bea} by 8/3 and 3, respectively
    (see \eq{eq:defb}).
  $m_B^{\rm pow}$ is a redundant parameter calibrating the overall size
  of the hadronic parameters $B_{R_i}$ which quantify the matrix
  elements at order $\lqcd/m_b$.  $B_{R_0}^q$ is calculated from 
  $\langle B_s| R_0|\bar{B}_s\rangle =- (0.66\pm 0.27) \gev^4$ 
  and
  $\langle B_d| R_0|\bar{B}_d\rangle= -(0.36 \pm 0.20) \gev^4$   
  \cite{Dowdall:2019bea} 
  (with $\langle R_0 \rangle$ defined as in
  Ref.~\cite{Beneke:1996gn,NiersteNLONB}) with the central values of $f_{B_q}$
  and the quark and meson masses listed above, so that the error quoted
  for $B_{R_0}^q$ correctly reflects the error of only the matrix
  element (and not the uncertainty of the artificial conversion factor from matrix elements
  to bag parameters). 
  In the same way  $B_{\tilde R_{2,3}}^s$ is calculated from  $\langle B_s| \tilde R_2
  |\bar{B}_s\rangle = (0.28 \pm 0.11) \gev^4  $ and
  $\langle B_s| \tilde R_3 |\bar{B}_s\rangle = (0.44 \pm 0.15) \gev^4  $
  \cite{Davies:2019gnp}.
  The expressions for $B_{R_2}^q$ and $B_{R_3}^q$
  hold up to $\lqcd/m_b$ corrections.  $B_{R_1}^q= 1.5$ and  $B_{\tilde
    R_1}^q= 1.2$ \cite{Davies:2019gnp} are phenomenologically irrelevant.
  The charm and bottom masses imply 
  $z={m_c^2(m_c)}/{m_b^2(m_b)}=0.096$ leading to
  $\bar{z}={m_c^2(m_b)}/{m_b^2(m_b)}=0.052 \pm 0.002$ at NLO
  and we use the same value at NNLO. \label{tab:inp}}
~\\[-3mm]\hrule
\end{table*}

For all  central values quoted in the following  we
took $\mu_1=m_b^{\rm pole}$ and $\mu_1=\bar{m_b}$ for the pole and
$\ov{\rm MS}$ schemes, respectively.
For the contribution to the width differences $\Delta\Gamma_s$ that
originates from the penguin sector and is proportional to
$\alpha_s^2N_f$ (neglecting $\lambda_u$ part) we find
\begin{eqnarray}
\label{delgam}
&& \frac{\delta\Delta\Gamma_s^{(2),N_f,p}(z)}{\delta\Delta\Gamma_s^{(2),N_f,p}(0)} = 1.14 .
\end{eqnarray}
\eq{delgam} shows that the effect of a non-zero charm quark mass on the
lines touching $O_2$ are important for the penguin contribution, leading
to an about $14\%$ increase of the $\alpha_s^2N_f$ contribution to the
latter in comparison to the case in which the charm quark mass on all
lines touching $O_2$ is set to zero.

The penguin contribution at order $\alpha_s$ \cite{NiersteNLO} evaluates  to 
\begin{eqnarray}
\label{delgam2}
  \nonumber \frac{\delta\Delta\Gamma_s^{(1),p}(z)}{\Delta\Gamma_s^{\rm
  NLO}(z)}
  &=& -14.5\%~~~(\rm{pole}),
  \\
  \frac{\delta\Delta\Gamma_s^{(1),p}(z)}{\Delta\Gamma_s^{\rm NLO}(z)}
  &=& -11.2\%~~~(\ov{\rm{MS}}),
\end{eqnarray}
and the new $\alpha_s^2N_f$ corrections are
\begin{eqnarray}
%\label{delgam3}
\nonumber \frac{\delta\Delta\Gamma_s^{(2),N_f,p}(z)}{\Delta\Gamma_s^{\rm NLO}(z)} &=& 2.4\%,
~~~(\rm{pole}), \\
\frac{\delta\Delta\Gamma_s^{(2),N_f,p}(z)}{\Delta\Gamma_s^{\rm NLO}(z)} &=& 1.8\%,
                                                                            ~~~(\ov{\rm{MS}}),
                                                                            \label{dgp2num}                                                          
\end{eqnarray}
where $\delta\Delta\Gamma_s^{(1),p}(z)$ denotes the contribution to
$\Delta\Gamma_s$ from the penguin sector at order $\alpha_s$ and
$\delta \Delta\Gamma_s^{(2),N_f,p}(z)$ is the corresponding contribution 
at order $\alpha_s^{2}N_f$.

The analogous contributions to the  CP asymmetries
at NLO \cite{Beneke:2003az}  are
\begin{eqnarray}
\nonumber \frac{ \delta a_{\rm fs}^{q(1),p}}{ a_{\rm fs}^{s,\rm NLO}} &=& 3.0\% ,~~~(\rm{pole}), 
\\
\frac{ \delta a_{\rm fs}^{q(1),p}}{ a_{\rm fs}^{s,\rm NLO}} &=& 2.7\% ,~~~(\ov{\rm{MS}}),
\end{eqnarray}
with $q=s,d$, while at order $\alpha_s^2N_f$ we obtain
\begin{eqnarray}
\nonumber \frac{ \delta a_{\rm fs}^{q(2),N_f,p}}{ a_{\rm fs}^{q,\rm NLO}} &=& -1.2\%~~~(\rm{pole}), 
\\
  \frac{ \delta a_{\rm fs}^{q(2),N_f,p}}{ a_{\rm fs}^{q,\rm NLO}} &=& -1.0\% .~~~(\ov{\rm{MS}}),
\label{afsp2num}                                                         
\end{eqnarray}
Judging from the numbers presented above, we see that the penguin
contributions at order $\alpha_s^2N_f$ have opposite sign compared to
the ${\cal O}(\alpha_s)$ penguin corrections and decrease the latter by
approximately 37\%. This nurtures the expectation that the full
$\alpha_s^2$ corrections may also be large and a reliable assessment of
the penguin contribution calls for a complete NNLO calculation. For the SM
contribution considered here the overall contributions to $\dg_q$ and
$a_{\rm fs}^q$ is small (see \eqsand{dgp2num}{afsp2num}), but in BSM
models with enhanced penguin coefficients these corrections are relevant
to constrain these coefficients from the data.

Until the full NNLO calculation is available, we recommend to use the
following updated NLO SM values for $\dg_q/\dm_q$:
\begin{eqnarray}
  \nonumber \frac{\dg_s}{\dm_s}
   &=&(4.33 \pm 0.83_{\rm{scale}}\pm 0.11_{B,\widetilde{B}_S}\pm 0.94_{\lqcd/{m_b}})
\\
&& \nonumber \times 10^{-3} ~~(\rm{pole}),
  \\ \nonumber
  \frac{\dg_s}{\dm_s}
  &=& (4.97 \pm 0.62_{\rm{scale}}\pm 0.13_{B,\widetilde{B}_S}\pm 0.80_{\lqcd/{m_b}})
\\
&& \times 10^{-3} ~~ (\ov{\rm MS}), \label{degabsnum} \\
%\end{eqnarray}
%\begin{eqnarray}
  \nonumber \frac{\dg_d}{\dm_d}
   &=&(4.48 \pm 0.82_{\rm{scale}}\pm 0.12_{B,\widetilde{B}_S}\pm 0.86_{\lqcd/{m_b}})
\\
&& \nonumber \times 10^{-3} ~~(\rm{pole}),
  \\ \nonumber
  \frac{\dg_d}{\dm_d}
    &=&(5.07 \pm 0.61_{\rm{scale}}\pm 0.14_{B,\widetilde{B}_S}\pm 0.73_{\lqcd/{m_b}})
\\
&& \times 10^{-3} ~~ (\ov{\rm MS}) \label{degabdnum}
\end{eqnarray}
and $a_{\rm fs}^q$:
\begin{eqnarray}
\label{afss}
\nonumber a_{\rm fs}^{s}&=&(2.07\pm 0.08_{\rm{scale}}\pm 0.02_{B,\widetilde{B}_S}\pm 0.05_{\lqcd/{m_b}}
\\
&& \nonumber  \quad \pm 0.04_{\rm CKM}) \times 10^{-5} ~~(\rm{pole}),
\\
\nonumber a_{\rm fs}^{s}&=&(2.04\pm 0.09_{\rm{scale}}\pm 0.02_{B,\widetilde{B}_S}\pm 0.04_{\lqcd/{m_b}}
\\
&& \quad \pm 0.04_{\rm CKM}) \times 10^{-5} ~~ (\ov{\rm MS}),
\end{eqnarray}
\begin{eqnarray}
\label{afsd}
\nonumber a_{\rm fs}^{d}&=&-(4.71\pm 0.18_{\rm{scale}}\pm 0.04_{B,\widetilde{B}_S}\pm 0.11_{\lqcd/{m_b}}
\\
&& \nonumber \qquad \pm 0.10_{\rm CKM}) \times 10^{-4} ~~(\rm{pole}),
\\
\nonumber a_{\rm fs}^{d}&=&-(4.64\pm 0.21_{\rm{scale}}\pm 0.04_{B,\widetilde{B}_S}\pm 0.09_{\lqcd/{m_b}}
\\
&& \qquad \pm 0.10_{\rm CKM}) \times 10^{-4} ~~(\ov{\rm MS}).
\end{eqnarray}
The error indicated with ``$\lqcd/{m_b}$'' comprises the uncertainty
from the bag factors of Refs.~\cite{Dowdall:2019bea,Davies:2019gnp}. The
new lattice results for the bag parameters of the $\lqcd/{m_b}$
corrections have errors comparable to those assumed in
Ref.~\cite{Asatrian:2017qaz}, but the central value of ${B}_{R_0}^s$ has
shifted upwards by more than a factor of 2. Furthermore,
$\widetilde{B}_{S,B_s}^\prime/B_{B_s}$ decreased by 12\%, which also
lowered the $\mu_1$ dependence. Adding the individual errors quoted in
\eqsto{degabsnum}{afsd} in quadrature yields the values quoted in the
abstract.

With the input values of \tab{tab:inp} we reproduce the measured $\dm_s$
in an excellent way.  It makes therefore no difference, whether we use
the experimental or theoretical value to calculate $\dg_s$ from the
ratios in \eq{degabsnum}.  The central values for $\dg_s$ in
Ref.~\cite{Asatrian:2017qaz} are proportional to $B_{B_s}$, and the
value used in that analysis was larger than the one in \tab{tab:inp} by
16\%, explaining why the $\dg_s$ values in Ref.~\cite{Asatrian:2017qaz}
were larger by roughly the same amount compared to
\begin{align}
     \dg_s^{\rm pole}&=(0.077 \pm 0.022)\, \mbox{ps}^{-1}, \nn
     \dg_s^{\overline{\rm MS}}&=(0.088\pm 0.018)\, \mbox{ps}^{-1} 
\end{align}
inferred from \eq{degabsnum} with
$\dm_s^{\rm exp} = (17.757 \pm 0.021)\, \mbox{ps}^{-1}$.

In $a_{\rm fs}^q$, however, the lattice results for the $\lqcd/{m_b}$
bag parameters already have an impact on reducing the uncertainty,
because unlike $\dg_q/\dm_q$ the CP asymmetry $a_{\rm fs}^q$ is very
sensitive to ${B}_{\tilde R_3}^s$, whose uncertainty of $\pm 0.39$ is
below the $\pm 0.5$ assumed in older analyses done without the lattice
input.

The scale dependence is calculated by varying $\mu_1$ between
$m_b/2$ and $2m_b$. Both this scale dependence and the
sizable scheme dependence indicate  that the 
missing perturbative higher-order corrections in
$\dg_q/\dm_q$ are not small. However, the  $\mu_1$ dependence
might well underestimate this error in the case of $a_{\rm fs}^q$.

The central values of all our $\overline{\rm MS}$ scheme results are in
excellent agreement with Ref.~\cite{Lenz:2019lvd}. Our error estimate of
the $\lqcd/{m_b}$ corrections is conservative, as we add the errors of
individual bag parameters linearly,  leading to overall uncertainties in
$\dg_q/\dm_q$ which are larger by roughly a factor of 1.5 compared to
those of $\dg_q$ in Ref.~\cite{Lenz:2019lvd}. Our uncertainties for
$a_{\rm fs}^q$, though, are smaller compared to
Ref.~\cite{Lenz:2019lvd}, as we find a smaller $\mu_1$-dependence
and assume a smaller error on $m_c$. (Recall that
$a_{\rm fs}^q\propto m_c^2$.) In our error budget the 0.9\% error in
$m_c$ quoted in \tab{tab:inp} would contribute another 3\% uncertainty
to $a_{\rm fs}^q$.\\

\section{Conclusions}
We have calculated the penguin contributions of order $\alpha_s^2 N_f$
to the width difference $\dg_q$ and the CP asymmetry in flavor-specific
decays of $B_q$ mesons, $a_{\rm fs}^q$. These and the mass difference $\dm_q$
are fundamental quantities characterizing the \bbmq\
complex.  The calculation improves over Ref.~\cite{Asatrian:2017qaz} by
taking into account the full dependence on the charm quark mass.  In
line with the general findings of Ref.~\cite{Beneke:2002rj} we find no
enhancement proportional to $\log(m_b^2/m_c^2)$ in the new terms of
order $\alpha_s^2 N_f m_c^2/m_b^2$, but we discover a largish
coefficient of this term and conclude that the future full NNLO
calculation of the penguin pieces should incorporate the full
$m_c$-dependence.  In both $\dg_q$ and $a_{\rm fs}^q$ the
$\alpha_s^2 N_f$ terms have signs opposite to the NLO corrections. The
calculated partial NNLO corrections are smaller than the corresponding
NLO terms by factors of roughly 6 and 3 for $\dg_s$ and $a_{\rm fs}^q$,
respectively, indicating a good convergence of the perturbative series.

In response to the recent progress in the lattice calculations of the
non-perturbative matrix elements \cite{Dowdall:2019bea,Davies:2019gnp} 
we have further presented updated NLO values for
$\dg_q$ and $a_{\rm fs}^q$ in Eqs. (\ref{degabsnum}) to (\ref{afsd}).

\begin{acknowledgments}
  This work has been supported by Grant No. 86426 of
  VolkswagenStiftung.
  H.A., A.H. and A.Y.\ have further been supported by the
  State Committee of Science of Armenia Program Grant No. 18T-1C162, and
  S.T.\ was supported within the Regional Doctoral Program on
  Theoretical and Experimental Particle Physics sponsored by
  VolkswagenStiftung.
  U.N.\ is supported by BMBF under grant
  \emph{Verbundprojekt 05H2018 (ErUM-FSP T09) - BELLE II: Theoretische
    Studien zur Flavourphysik} and by project C1b of the DFG-funded
  Collaborative Research Center TRR 257, ``Particle Physics
  Phenomenology after the Higgs Discovery''.
\end{acknowledgments}

\appendix
\section{Full-theory matrix elements\label{appa}}

In this section we collect the needed unrenormalized LO and NLO matrix
elements to order $\epsilon^2$ and $\epsilon$, respectively, where
$\epsilon=(4-D)/2$ appears in the ultraviolet poles in dimensional
regularization.  We decompose the matrix element as
\begin{equation}
M \,= \, M_{\rm cc} + M_{\rm peng}, \label{mccp}
\end{equation}
where the first term denotes the contribution with two insertions
of the current-current operators $O_{1,2}$ and the second term
comprises the diagrams with at least one penguin operator. Recall that
we count $C_{3-6}$ as order $\alpha_s$, so that one loop less is needed
for $M_{\rm peng}$ compared to $M_{\rm cc}$. We expand
$M_{\rm peng}^{ab}=M_{\rm peng}^{ab,(0)} + \frac{\alpha_s}{4\pi}M_{\rm peng}^{ab,(1)}+\ldots$.

\begin{widetext}
\subsection{Penguin operators}

Here and in the following $\langle ...\rangle^{(0)}$ denotes tree-level matrix element and
$C_k^b=\sum_j C_j Z_{jk}$ are bare Wilson coefficients (see eq. (3.10) of \cite{Asatrian:2017qaz}).

We decompose the NLO penguin diagrams according to the diagrams in figure \ref{diags} as
\begin{eqnarray}
% && \nonumber M^{(1)}_{\rm peng} = -\frac{G^2_F m_b^2}{12\pi} \left[ \lambda_c^2 \left( M^{cc,(1)}_{D_{11}} + M^{c,(1)}_{D_{12}} \rt) \right.
% \\
% && \nonumber + \lambda_c\lambda_u \left(2M^{cu,(1)}_{D_{11}} + M^{c,(1)}_{D_{12}} + M^{u,(1)}_{D_{12}} \right) 
% \\
% && \left. + \lambda_u^2 \lt( M^{uu,(1)}_{D_{11}} + M^{u,(1)}_{D_{12}} \right) \right],
\nonumber M^{(1)}_{\rm peng} &=& -\frac{G^2_F m_b^2}{12\pi} \left[
                                 \lambda_c^2
                                 \left( M^{cc,(1)}_{D_{11}} + M^{c,(1)}_{D_{12}} \rt) 
   + \lambda_c\lambda_u \left(2M^{cu,(1)}_{D_{11}} + M^{c,(1)}_{D_{12}} + M^{u,(1)}_{D_{12}} \right) 
   + \lambda_u^2 \lt( M^{uu,(1)}_{D_{11}} + M^{u,(1)}_{D_{12}} \right) \right],
\end{eqnarray}
where
\begin{eqnarray}
&& \nonumber M^{q_1q_2,(1)}_{D_{11}} =
-\frac{5\langle Q\rangle^{(0)}+8\langle \tilde{Q}_S\rangle^{(0)}}{9}
  C_{2}^{b2} \left(\frac{\sqrt{1-4z_1} (1+2z_1)+\sqrt{1-4z_2} (1+2z_2)}{2\epsilon} \right.
\\
&& \nonumber \hspace{0.0cm} + \sqrt{1-4 z_2} \left(\frac{1}{2} (2 z_2+1) \left(4 \log \left(\frac{\mu_1}{m_b}\right)-\log (z_1)-\log (1-4 z_2)\right)+2 z_1  (2 z_2+1)+\frac{1}{3} (7 z_2+2)\right)
\\
&& \nonumber \hspace{0.0cm} +\sqrt{1-4 z_1} \left(\frac{1}{2} (2 z_1+1) \left(4 \log \left(\frac{\mu_1}{m_b}\right)-\log
   (1-4 z_1)-\log (z_2)\right)+2 (2 z_1+1) z_2+\frac{1}{3} (7 z_1+2)\right)
\\
&& \nonumber \hspace{0cm} +\frac{1}{2} \sqrt{1-4 z_1} (2 z_1+1) \sqrt{1-4
   z_2} (2 z_2+1) (\log (\sigma_1)+\log (\sigma_2))
+ \epsilon\left(\sqrt{1-4 z_1} \left(\left(\frac{2+7 z_1}{3}+2 (1+2 z_1) z_2\right)\right.\right.
\\
&& \nonumber \hspace{0cm} \times \left(4 \log \left(\frac{\mu_1}{m_b}\right)-\log (1-4z_1)-\log (z_2)\right)
 +\frac{1}{4} (1+2 z_1) \left(\left(4 \log \left(\frac{\mu_1}{m_b}\right)-\log (1-4 z_1)-\log (z_2)\right)^2-\frac{\pi ^2}{3}\right)
\\
&& \nonumber \left. + \frac{1}{3} (40 z_1 z_2+17 z_1+14 z_2+5)\right)
+\sqrt{1-4 z_2} \left(\left(\frac{2+7 z_2}{3}+2 (1+2 z_2) z_1\right) \left(4 \log \left(\frac{\mu_1}{m_b}\right)-\log (1-4z_2)-\log (z_1)\right)\right.
\\
&& \nonumber \hspace{0cm} \left. +\frac{1}{4} (1+2 z_2) \left(\left(4 \log \left(\frac{\mu_1}{m_b}\right)-\log (1-4 z_2)-\log (z_1)\right)^2-\frac{\pi ^2}{3}\right)+\frac{1}{3} (40 z_1 z_2+17 z_2+14 z_1+5)\right)
\\
&& \nonumber + \sqrt{1-4 z_1} \sqrt{1-4 z_2} \left(\frac{1}{3} (20 z_1 z_2+7 (z_1+z_2)+2) (\log (\sigma_1)+\log (\sigma_2))- (2 z_1+1) (2 z_2+1) \left(-\frac{1}{2} (\log (\sigma_1)+\log (\sigma_2)) \right.\right.
\\
&& \left.\left.\left.\left. \times \left(4 \log \left(\frac{\mu_1}{m_b}\right)-\log (1-4 z_1)-\log (1-4 z_2)\right)  +\text{Li}_2(\sigma_1)+\text{Li}_2(\sigma_2)+\frac{1}{4} \left(\log ^2(\sigma_1)+\log ^2(\sigma_2)\right)+\frac{2 \pi ^2}{3}\right)\right) \right)\right),
\end{eqnarray}
\begin{eqnarray}
&& M^{q_1,(1)}_{D_{12}} =
-\frac{1}{3}\left(5\langle Q\rangle^{(0)}+8\langle \tilde{Q}_S\rangle^{(0)}\right)
 C_2^bC_8^b
 \\
 && \nonumber \hspace{3cm} \cdot\sqrt{1-4 z_1} \left(1+2z_1+\epsilon\left(\frac{2(1+5z_1)}{3}
+ (1+2z_1)\left(2\log \frac{\mu_1}{m_b}-\log(1-4z_1)\right)\right)\right).
\end{eqnarray}
$q_1$ and $q_1$ represent either $c$ or $u$ quark; and $z_1$ and $z_2$
are equal to $m_c^2/m_b^2$ when originating from the operator $O_2$ or equal
to zero when related to a $u$ quark associated with operator $O_2^u$.

For the matrix elements with one QCD penguin operator we write
\begin{equation}
  M_{\rm peng} = -\frac{G^2_F m_b^2}{12\pi} \,
  \sum_{j=3}^6 \lt[\lambda_c^2 M_{j2}(z) + \lambda_c\lambda_u
  \big( M_{j2}(z)+M_{j2}(0)\big) + \lambda_u^2 M_{j2}(0) \rt] .
\end{equation}
As usual we expand $M_{jk}$ as
$M_{jk}=M_{jk}^{(0)}+\frac{\alpha_s}{4\pi} M_{jk}^{(1)}+\ldots$.
The unrenormalized LO and NLO matrix elements necessary for the
renormalization of the penguin diagrams $D_{11}$ and $D_{12}$ are the
following:
\begin{eqnarray}
&& M_{32}^{(0)}(z) = 2C_2^bC_3^b T_3,\qquad M_{42}^{(0)}(z) =
 2C_2^bC_4^b T_4, \nn && M_{52}^{(0)}(z) =  2C_2^bC_5^b T_5,
\qquad M_{62}^{(0)}(z) =  2C_2^bC_6^b T_5, \label{eq:m2i}
\end{eqnarray}
\begin{eqnarray}
&&M_{42}^{(1)}(z) =  \,
 2C_2^bC_4^b (N_HT_1+N_VT_2+N_LT_2'), \qquad M_{62}^{(1)}(z) =
 \, 2C_2^bC_6^b (N_HT_1+N_VT_2+N_LT_2'), \label{eq:m2426}
\end{eqnarray}
where
\begin{eqnarray}
&& \nonumber T_1 = -\frac{1}{9}(8 \langle \tilde{Q}_S\rangle^{(0)} + 5 \langle Q\rangle^{(0)}) \sqrt{1-4z} \left[\frac{1+2z}{2 \epsilon
} +\frac{19+44z}{6} + \frac{1}{2} \left(4 \log \left(\frac{\mu_1}{m_b}\right) - \log(1-4z)- \sqrt{3} \pi
\right)\right.
\\
&& \nonumber \hspace{0.0cm} + \epsilon  \left(\frac{1}{4} (1+2 z) \left(\left(4 \log \left(\frac{\mu_1}{m_b}\right)-\log (1-4 z)\right)^2-\frac{\pi ^2}{3}
+2 \sqrt{3} \pi  \left(\frac{\log (3)}{2}-4 \log \left(\frac{\mu_1}{m_b}\right)+\log (1-4 z)\right.\right.\right.
\\
&& \nonumber \left.\left. -\frac{3 i \left(\text{Li}_2\left(\frac{1}{6}
   \left(3-i \sqrt{3}\right)\right)-\text{Li}_2\left(\frac{1}{6} \left(3+i \sqrt{3}\right)\right)\right)}{\pi }\right)\right)
+\frac{19+44 z}{6} \left(4 \log \left(\frac{\mu_1}{m_b}\right)-\log (1-4 z)\right) -\frac{\sqrt{3} \pi}{2}  (3+8 z)
\\
&& \left.\left. +\frac{57+158 z}{6}\right)\right],
\end{eqnarray}
\begin{eqnarray}
&& \nonumber T_2 = -\frac{1}{9}(8 \langle \tilde{Q}_S\rangle^{(0)} + 5 \langle Q\rangle^{(0)})
\left[\frac{\sqrt{1-4z} \left(1+2z\right)+\sqrt{1-4z_i}\left(1+2z_i\right)}{2\epsilon}\right.
+ \frac{1}{2}\sqrt{1-4z}(1+2z)\sqrt{1-4z_i}(1+2z_i)
\\
&& \nonumber \hspace{0cm} \times\left(\log \left(\sigma\right)+\log(\sigma_i)\right) + \frac{1}{6}\sqrt{1-4z_i}\left(7+20z_i+3(1+2z_i)
\left(4z+4\log\left(\frac{\mu_1}{m_b}\right)-\log \left(z\right)-\log(1-4z_i)\right)\right)
\\
&& \nonumber \hspace{0cm} + \frac{1}{6}\sqrt{1-4z}\left(7+20z+3(1+2z)
\left(4z_i+4\log\left(\frac{\mu_1}{m_b}\right)-\log \left(z_i\right)-\log(1-4z)\right)\right)
\\
&& \nonumber \hspace{0.0cm} + \epsilon \frac{1}{12} \left(\sqrt{1-4z_i}\left(2(17+40z+54 z_i+104z z_i)+2 (7+12z+20 z_i+24z z_i) \cdot\left(4 \log \left(\frac{\mu_1}{m_b}\right) -\log\left(z\right) \right.\right.\right.
\\
&& \nonumber \left.\left. -\log(1-4 z_i)\right)
+ 3 (1+2 z_i) \left(\left(4\log\left(\frac{\mu_1}{m_b}\right)-\log \left(z\right) -\log (1-4 z_i) \right)^2 - \pi^2(1+2z_i) \right)\right)
\\
&& \nonumber \hspace{0.0cm} + \sqrt{1-4z}\left(2(17+40z_i+54 z+104z z_i)+2 (7+12z_i+20 z+24z z_i)
\cdot\left(4 \log \left(\frac{\mu_1}{m_b}\right) -\log\left(z_i\right)-\log(1-4 z)\right)\right.
\\
&&  \nonumber \left. \hspace{0.0cm} + 3 (1+2 z) \left(\left(4\log\left(\frac{\mu_1}{m_b}\right)-\log \left(z_i\right) -\log (1-4 z) \right)^2 - \pi^2(1+2z) \right)\right)
\\
&& \nonumber \hspace{0.0cm} + \sqrt{1-4 z} \sqrt{1-4 z_i} \left((1+2 z) (1+2 z_i) \left(6 (\log (\sigma)+\log (\sigma_i)) \left(4 \log\left(\frac{\mu_1}{m_b}\right)-\log (1-4 z)-\log (1-4 z_i)\right)\right.\right.
\\
&& \left.\left.\left.\left. \hspace{0.0cm} +12 \text{Li}_2(\sigma)+12
 \text{Li}_2(\sigma_i)+3 \log^2(\sigma)+3 \log^2(\sigma_i)+8 \pi ^2\right) + 2 (7+20 z+20 z_i+52 z z_i) \left(\log (\sigma)+\log (\sigma_i)\right)\right)\right)\right],
\end{eqnarray}
and
\begin{eqnarray}
 && \nonumber T_3 \!\!=\!\! \sqrt{1-4z}\left[\langle Q\rangle^{(0)} \frac{1}{2}(1-4z)\left(1+\epsilon
\left(\frac{2}{3}+2\log \left(\frac{\mu_1}{m_b}\right)-\log(1-4z)\right)\right.\right.
\\
&& \left. \nonumber \hspace{0.0cm} +\epsilon ^2 \left(
\left(2\log \left(\frac{\mu_1}{m_b}\right)-\log(1-4z)\right)\left(\frac{2}{3}+\log \left(\frac{\mu_1}{m_b}\right)-\frac{1}{2}\log(1-4z)\right)-\frac{\pi^2}{4}+\frac{13}{9}\right)\right)
\\
&& \nonumber \left. \hspace{0cm} - \langle \tilde{Q}_S \rangle^{(0)} \left(1+2z+\epsilon
\left(\frac{2}{3}(1+5z)+(1+2z)\left(2\log \left(\frac{\mu_1}{m_b}\right)-\log(1-4z)\right)\right)+\epsilon ^2 \left(
\frac{2}{3}(1+5z)\right.\right.\right.
\\
&& \hspace{0.0cm} \left.\left.\left. \times \left(2\log \left(\frac{\mu_1}{m_b}\right)-\log(1-4z)\right)+\frac{1+2z}{2}\left(\left(2\log \left(\frac{\mu_1}{m_b}\right)-\log(1-4z)\right)^2-\frac{\pi^2}{2}\right)
+\frac{13+56z}{9}\right)\right) \right]\! ,
\end{eqnarray}
\begin{eqnarray}
&& \nonumber T_4 \!\!=\!\! \sqrt{1-4z}\left[\langle Q\rangle^{(0)} \left(1-z+\epsilon
\left(\frac{2+z}{3}+(1-z)\left(2\log \left(\frac{\mu_1}{m_b}\right)-\log(1-4z)\right)\right)\right.\right.
\\
&& \left. \nonumber \hspace{0.0cm} +\epsilon ^2 \left(
\frac{2+z}{3}\left(2\log \left(\frac{\mu_1}{m_b}\right)-\log(1-4z)\right) + \frac{1-z}{2}\left(\left(2\log \left(\frac{\mu_1}{m_b}\right)-\log(1-4z)\right)^2-\frac{\pi^2}{2}\right)+\frac{13+2z}{9}\right)\right)
\\
&& \nonumber \left. \hspace{0cm} + \langle \tilde{Q}_S \rangle^{(0)} \left(1+2z+\epsilon
\left(\frac{2(1+5z)}{3}+(1+2z)\left(2\log \left(\frac{\mu_1}{m_b}\right)-\log(1-4z)\right)\right)+\epsilon ^2 \left(
\frac{2(1+5z)}{3}\right.\right.\right.
\\
&& \left.\left.\left. \times \left(2\log \left(\frac{\mu_1}{m_b}\right)-\log(1-4z)\right)+\frac{1+2z}{2}\left(\left(2\log \left(\frac{\mu_1}{m_b}\right)-\log(1-4z)\right)^2-\frac{\pi^2}{2}\right) + \frac{13+56z}{9}\right)\right) \right],
\end{eqnarray}
\begin{eqnarray}
 \nonumber T_5 \!\!&=&\!\! \langle Q\rangle^{(0)} 3 z \sqrt{1-4 z} \left[1 + \epsilon \left(2 \log \left(\frac{\mu_1}{m_b}\right)-\log (1-4 z)+1\right)\right.
\\
&& \left. \hspace{0cm} + \epsilon ^2 \left(\frac{1}{2} \left(2 \log \left(\frac{\mu_1}{m_b}\right)-\log (1-4 z)\right) \left(2 \log \left(\frac{\mu_1}{m_b}\right)-\log (1-4 z)+2\right)-\frac{\pi^2}{4}+2\right)\right],
\end{eqnarray}
where $z=m_c^2/m_b^2$ contains the charm mass on lines attached to $O_{1,2}$
and $z_i=m_c^2/m_b^2$ contains the charm mass from the
closed fermion loop. $T_2^\prime$ is obtained from $T_2$ by setting
$z_i$ to zero. For the matrix elements with two QCD penguin operators we
refer to Eqs.~(A.15)-(A.18) in Ref.~\cite{Asatrian:2017qaz}.

\end{widetext}

\end{document}